\documentclass[aps,twocolumn,prx,amsfonts,showpacs,superscriptaddress,longbibliography]{revtex4-2}
\usepackage{graphicx}
\usepackage{bm}
\usepackage{hyperref}
\usepackage{amsmath}
\usepackage{amssymb}
\usepackage[table]{xcolor}
\usepackage{array}
\usepackage{multirow}
\usepackage[caption=false]{subfig}
\usepackage{physics}
\usepackage{booktabs}
\usepackage{soul}
\usepackage{amsthm}
\usepackage{bbm}
\usepackage{multirow}
\usepackage[scr=boondoxo,frak=boondox,scrscaled=1.05]{mathalfa}
\usepackage{amsmath}

\def\bra#1{\langle#1|}
\def\ket#1{|#1\rangle}
\def\braket#1#2{\langle#1|#2\rangle}

\def\r{{\boldsymbol{r}}}
\def\x{{\boldsymbol{x}}}
\def\k{{\boldsymbol{k}}}
\def\p{{\boldsymbol{p}}}
\def\q{{\boldsymbol{q}}}

\def\g{{\boldsymbol{g}}}

\def\R{{\boldsymbol{R}}}
\def\G{{\boldsymbol{G}}}

\def\X{{\boldsymbol{X}}}
\def\a{{\boldsymbol{a}}}
\def\b{{\boldsymbol{b}}}
\def\L{{\boldsymbol{L}}}

\newcommand{\br}{\bm {r}}

\usepackage{ulem}
\begin{document}

\title{Higher-Winding Fractionalization}

\author{Kishore Iyer}

\affiliation{Universit\'e Paris Cit\'e, CNRS,  Laboratoire  Mat\'eriaux  et  Ph\'enom\`enes  Quantiques, 75013  Paris,  France}

\author{Christophe Mora}

\affiliation{Universit\'e Paris Cit\'e, CNRS,  Laboratoire  Mat\'eriaux  et  Ph\'enom\`enes  Quantiques, 75013  Paris,  France}

\author{Daniele Guerci}

\affiliation{Department of Physics, William \& Mary, Williamsburg, VA-23187, USA}
\affiliation{Department of Physics, Massachusetts Institute of Technology, Cambridge, MA-02139, USA}

\begin{abstract}

Higher-winding skyrmion textures can generate emergent magnetic fields with multiple flux quanta per unit cell.
This opens an intriguing route toward fractionalization, allowing fractionalized quantum anomalous Hall states to arise even at integer filling of the microscopic unit cell. We show that, in this setting, increasing lattice-scale inhomogeneity of the emergent magnetic field drives a Berezinskii--Kosterlitz--Thouless (BKT) transition between a fractionalized liquid and a crystalline dielectric state. 
This transition carries a topological signature: under flux insertion, the many-body polarization defines a quantized winding number that is nonzero in the fractionalized phase and vanishes in the dielectric crystal.

\end{abstract}
\maketitle
\date{\today}

\textit{Introduction ---} Internal degrees of freedom, such as spin~\cite{Bruno2004,Hamamoto2015,Paul_2023} or layers in moir\'e heterostructures~\cite{FW_PRL_2019,Dong2023,MoralesDuran2024,Shi2024}, have emerged as a powerful route to engineer emergent magnetic fields~\cite{kolar2026}. 
This mechanism underpins the recent breakthrough realizations of the fractional quantum anomalous Hall effect in twisted transition-metal dichalcogenides ($t$TMDs)~\cite{Cai2023,zeng2023thermodynamic, xuParkObservationFractionallyQuantized2023,PhysRevX.13.031037}. A similar picture has been considered for rhombohedral graphene~\cite{tan2025ideallimitrhombohedralgraphene,desrochers2026energeticsfractionalanomaloushall,maymann2026skyrmionfractionalcherninsulator}, where the fractional Chern insulator (FCI)  has been observed in the presence of aligned hBN~\cite{Lu2024Feb,lu_extended_2025,xie2025tunablefractionalcherninsulators,Choi2025,Aronson2025}.
Beyond solid-state systems, this idea is also relevant to cold-atoms where emergent magnetic fields can be realized in optical flux lattices~\cite{L_onard_2023,kwan2026pfaffianquantumhallstate} by nontrivial winding of internal atomic states~\cite{Cooper2011,Cooper2012,Cooper2013,sommer2025idealopticalfluxlattices}.

Emergent magnetic fields provide a versatile platform for realizing fractionalized topological phases beyond conventional quantum Hall analogs.
Lattice-scale modulation breaks continuous magnetic translation symmetry, endowing charged excitations with a finite dispersion~\cite{Abouelkomsan2020,Regnault2026,yan2025anyondispersionaharonovcasherbands,schleith2025anyondispersionnonuniformmagnetic,iyer2026dispersionanyonblochbands}. 
This enables unconventional correlated phenomena, most notably superconductivity proximate to FCIs~\cite{DGAA2025,guerci2026topologicalsuperconductivityemergentvortex,wang2025chiralsuperconductivitynearfractional}, as suggested by recent observations in $t$TMDs~\cite{xu2026signaturesunconventionalsuperconductivitynear}, as well as optical responses~\cite{paul2025shininglightcollectivemodes,Kousa2025} forbidden in the uniform-field limit~\cite{Kohn1961}. At the same time, the modulation can either enhance~\cite{sarkar2026similarfractionalcherninsulators} or suppress~\cite{Shi2026,morales2026bandmixingparticleholeasymmetry,he2025fractionalcherninsulatorscompeting} the stability of fractionalized phases. Two distinct routes extend this landscape: engineering bands with higher Chern number~\cite{Wu2013,Wilhelm2023,dong2023_color,Wang2023,GuerciPRB2025}, or exploiting higher-winding skyrmion textures carrying multiple flux quanta per unit cell~\cite{Cooper2011,Sarkar2025}.

Here, we explore the second route and investigate the interacting phases arising in skyrmion lattices with higher winding number $Q>1$. In these systems, the skyrmion texture generates an emergent magnetic field carrying $Q$ flux quanta, $Qh/e$, per unit cell, allowing the emergent flux and the electronic filling $\nu$ of the underlying unit cell to be varied independently. 
We show that this enables fractionalization at integer filling of the unit cell and hence, for a fixed lattice period, at higher electron densities.
More fundamentally, higher-winding skyrmion textures decouple microscopic filling from many-body topology, enlarging the space of accessible fractionalized phases.

\begin{figure}
    \centering
    \includegraphics[width=\linewidth]{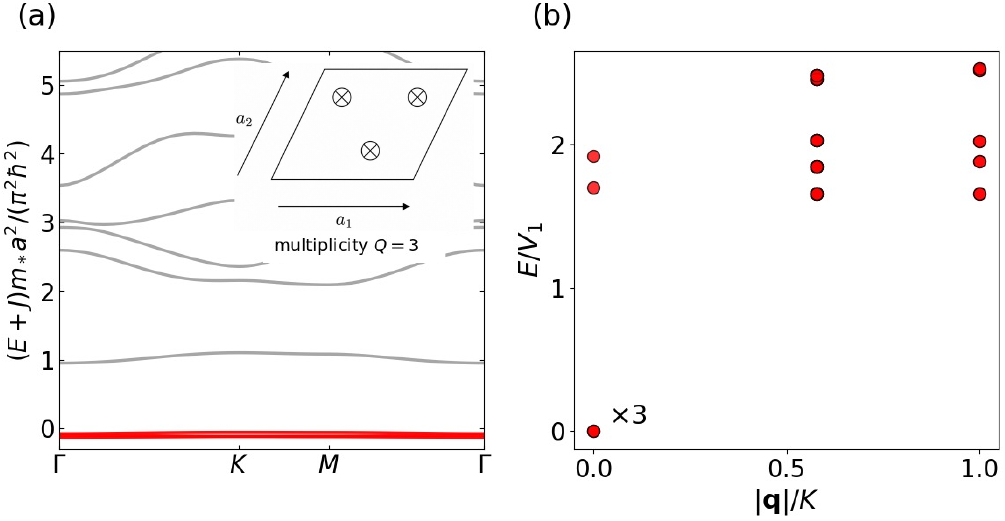}
    \caption{\textbf{Higher-winding skyrmion model bands and fractionalization at filling $\nu=1$}. 
    The low-energy manifold for $Jm_* a^2/(\pi^2\hbar^2)=100$ consists of $Q=3$ flat bands with total Chern number $|C_{\rm tot}|=1$, which become degenerate in the infinite-$J$ limit. 
    Many-body spectrum vs. center of mass momentum $\q$, from ED on a $3\times3$ torus for electrons at filling $\nu=1$, using the three lowest energy bands with $V_1$ interaction at $\mathcal{K}=0.4$. 
    The ground state is threefold degenerate.}
\label{fig:bandsskyrmion}
\end{figure}

We focus on the minimal higher-winding case, $Q=3$, at filling $\nu=1$ of the ideal bands. 
In this regime, lattice-scale inhomogeneities of the emergent magnetic field qualitatively reshape the many-body ground state.
For weak modulation, repulsive interactions stabilize a FCI with a threefold-degenerate ground-state manifold adiabatically connected to the Laughlin fractional quantum Hall state.
Interestingly, increasing the inhomogeneity of the emergent magnetic field drives a transition to a topologically trivial state in which the electrons are pinned at the minima of the emergent magnetic field, even for short-range interactions. 
By mapping the many-body groundstate probability distribution onto a two-dimensional Coulomb gas, we identify the two phases as a screening plasma and a dielectric of confined particle-hole dipoles, respectively, and derive the Berezinskii--Kosterlitz--Thouless (BKT) renormalization-group flow governing the transition. 
Closely related plasma-mapping results have recently been obtained for ideal Chern bands~\cite{Moitra2026}; see also Ref.~\cite{JC_trithep}. Crucially, this transition is not a generic consequence of field inhomogeneity. The $Q=1$ model at filling $1/3$ remains a screening liquid even at strong modulation, identifying the integer commensurability of the $Q=3$ model as an essential ingredient for the dielectric instability.

We complement the analytical results with extensive numerical calculations, combining exact diagonalization (ED) and large-scale quantum Monte Carlo simulations. We track the liquid-to-dielectric transition through the many-body spectrum, the long-wavelength static structure factor and the many-body polarization. 
Remarkably, we find that the BKT transition carries a topological signature in the many-body polarization: its quantized winding number under flux insertion is nonzero in the FCI phase and vanishes in the dielectric phase. The center-of-mass fluctuations develop a pronounced peak precisely as the polarization ceases to wind, signaling the loss of Hall response and the onset of the dielectric phase.

\textit{Higher-Winding Skyrmion Bands ---} We begin by examining the electronic bands induced by strong exchange coupling to a higher-winding skyrmion texture. 
To this aim, we consider a general Hamiltonian for electrons in low-density systems relevant to twisted TMDs~\cite{FW_PRL_2019} and to two-dimensional electrons coupled to magnetic textures~\cite{Bruno2004,Hamamoto2015,Paul_2023}:
\begin{equation}\label{skyrmionmodel}
    H(\r)=-\frac{\hbar^2\nabla^2}{2m_*}+ J\bm n(\r)\cdot \bm \sigma + V(\r),
\end{equation}
where $m_*$ is the effective mass, $J$ denotes the pseudospin exchange coupling to the texture, $\bm n(\r)$ is a unit vector describing a skyrmion texture, $\boldsymbol{\sigma}$ are the Pauli matrices acting on the internal degree of freedom.
Both $\bm n(\r)$ and the moiré potential $V(\r)$ share the periodicity of the lattice, defined by primitive vectors $\a_1$ and $\a_2$ of equal length $a$ and fixed relative angle, with unit-cell area $A_{\rm UC}=|\a_1\times\a_2|$. 
This discussion generalizes to multicomponent systems with more than two internal degrees of freedom, such as trilayer $t$TMDs~\cite{Choi_2025,Nakatsuji_2025}, by replacing the Pauli matrices with the  generators of the corresponding  group~\cite{Graf2021}.

In the large-exchange limit, $Jm_* a^2/\hbar^2\gg1$, the spinor component of the Bloch wavefunction  follows the local texture adiabatically. 
Within the adiabatic approximation~\cite{MoralesDuran2024}, we transform the Hamiltonian to a basis where the spinor is locally aligned with  $\bm n(\r)$. 
As shown in~\cite{supplementary}, in the $J\to\infty$ limit, a suitable choice of $V(\br)$ yields the exact Aharonov--Casher Hamiltonian~\cite{Shi2024},
\begin{equation}\label{eq:AC}
        H_{\rm AC} = \frac{\hbar^2}{2m_*}\left[-i\partial+\mathcal A^*(\r)\right]\left[-i\bar\partial+\mathcal A(\br)\right],
\end{equation}
with the notations $\partial = \partial_x - i \partial_y$ and $\mathcal A = \mathcal A_x + i \mathcal A_y$. 
Eq.~\eqref{eq:AC} corresponds to the square of a Dirac operator~\cite{dong2022diracelectronperiodicmagnetic} coupled to the emergent vector potential $\mathcal A$.
The associated lattice-periodic magnetic field $\mathcal B(\br) = -\, \bm n(\br) \cdot [ \partial_x \bm n(\br) \times \partial_y \bm n(\br)]/2$ reflects the non-trivial winding of the skyrmion texture.
For the sign of the flux considered here, the zero modes of Eq.~\eqref{eq:AC} lie in the holomorphic sector.
The corresponding number of emergent flux quanta per unit cell $n_\Phi$ is quantized, equal and opposite to the Pontryagin index of the skyrmion $Q$,
\begin{equation}\label{eq:net_flux}
Q=-n_{\Phi}=-\int_{\rm UC}\frac{d^2\r}{2\pi} \mathcal B(\r) \in \mathbb{N}.
\end{equation}
We decompose the magnetic field as $\mathcal B(\br) = \mathcal B_{n_\Phi} + \delta \mathcal B(\br)$, where the uniform component threads a flux $A_{\rm UC}\mathcal B_{n_\Phi}=2\pi n_\Phi$ through each unit cell, while the periodic modulation $\delta\mathcal B(\br)=-\nabla^2K(\br)$ has zero net flux over the unit cell with $K(\r)$ being the Kähler potential.

The Aharonov–Casher Hamiltonian in Eq.~\eqref{eq:AC} admits exact zero-energy solutions, with their number per unit cell equal to the topological charge $Q$ of the skyrmion texture~\eqref{eq:net_flux}, as sketched in the inset of Fig.~\ref{fig:bandsskyrmion}(a).
This result directly carries over to the original Hamiltonian in Eq.~\eqref{skyrmionmodel} in the $J\to\infty$ limit, where the spinor is rigidly locked to the local skyrmion texture.
The corresponding orthonormal zero-energy eigenstates read
\begin{equation}\label{wavefunction}
    \vec \psi_{\k n}(\r)=\vec \chi(\r)\, e^{- K(\r)} \varphi_{\k n}(\r), 
\end{equation}
where $\varphi_{\k n}(\r) $ are quasiperiodic lowest Landau levels and $\vec \chi(\r)$ is a normalized spinor, whose real space winding screens the net magnetic field carried by the lowest Landau level wavefunctions~\cite{GuerciPRB2025}. At each momentum $\k$, the label $n$ takes $Q$ distinct values. 
This multiplicity, dictated by the Atiyah--Singer index theorem~\cite{atiyah1968index,AtiyahSinger1984}, can be understood through band folding. 
Specifically, the system may be viewed in terms of a reduced unit cell of area $A_{\rm UC}/Q$, pierced by a single flux quantum, and therefore a $Q$-fold enlarged Brillouin zone. Bloch--Landau levels defined in this enlarged Brillouin zone carry Chern number $|C|=1$. 
Folding it into the original Brillouin zone maps $Q$ distinct states onto each momentum $\k$ and produces an $Q$-fold degenerate manifold. By conservation of the topological index under band folding, this manifold retains the total Chern number $|C_{\rm tot}|=1$.

We construct square- and triangular-lattice models that realize and verify this picture, using a skyrmion texture with real-space winding number $Q=3$ per unit cell~\cite{supplementary}.
For sufficiently strong exchange coupling $J$ and a moiré potential $V(\br)$ tuned to satisfy the Aharonov--Casher mapping, we numerically solve Eq.~\eqref{skyrmionmodel}. 
The resulting spectrum, shown in Fig.~\ref{fig:bandsskyrmion}(a), exhibits three nearly degenerate flat bands near zero energy, separated from higher-energy states by a spectral gap. 
In agreement with the band-folding construction, these three bands carry total Chern number $|C_{\rm tot}|=1$.

\begin{figure}
    \centering
    \includegraphics[width=\linewidth]{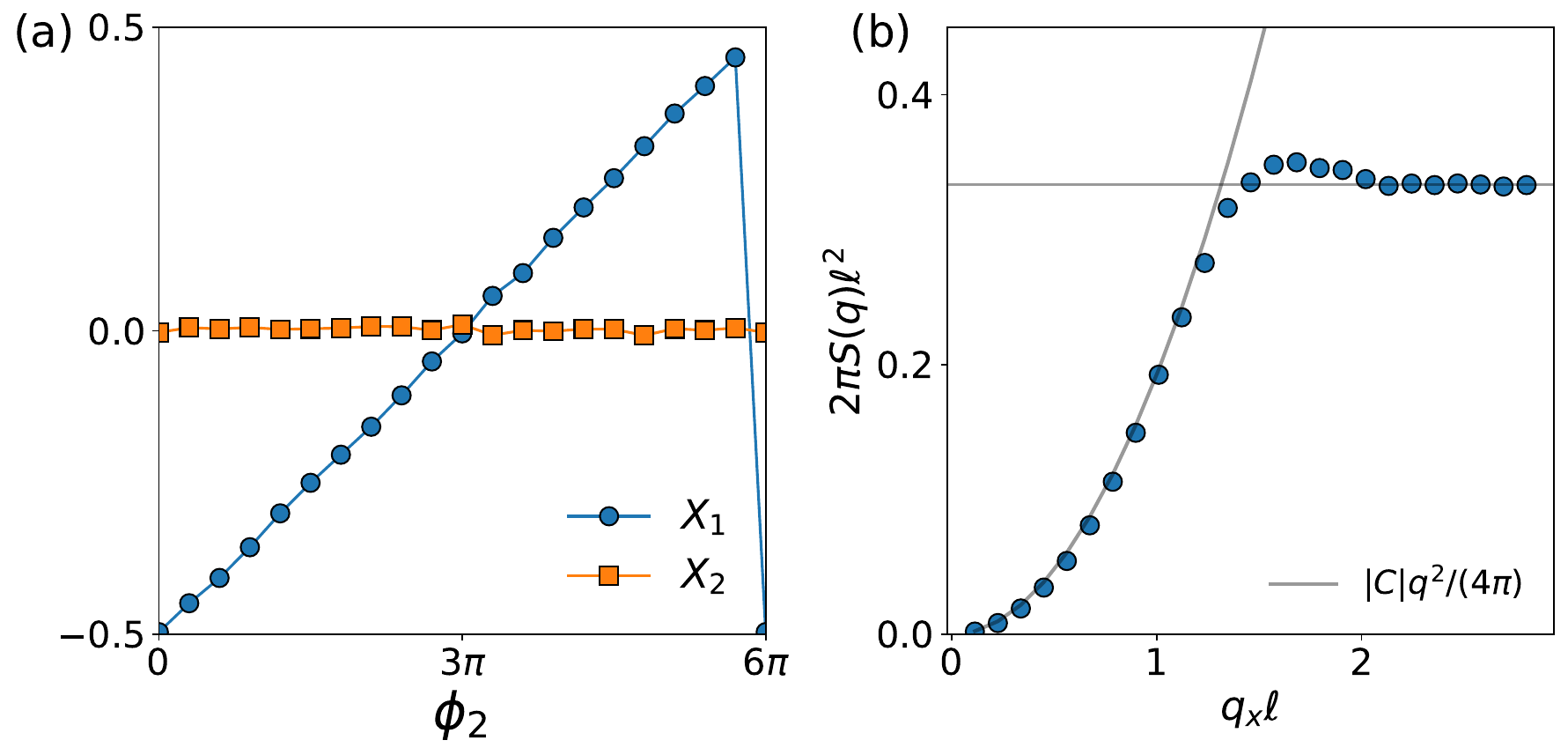}
    \caption{\textbf{Topological response and perfect screening of the ideal FCI}. 
    (a) Many-body positions $X_1$ (blue) and $X_2$ (orange) under flux insertion $\phi_2$.
    (b) Connected static structure factor $S(\q)$ along $q_x$ in unit of $\ell^2=A_{\rm UC}/(2\pi Q)$, where the solid parabola shows the long-wavelength behavior. 
    Plots are for $N=192$ electrons at $\mathcal K=0.4$.}
    \label{fig:comflux_structurefactor}
\end{figure}

\textit{Fractionalization at integer filling ---} Having established the structure of the single-particle flat-band manifold~\eqref{wavefunction}, we now turn to the many-body problem obtained by partially filling it.  
The crucial point is that, for $Q>1$, an integer filling of the microscopic unit cell, $\nu=n_eA_{\rm UC}$, corresponds to a fractional filling of the $Q$-fold degenerate manifold, where $n_e$ is the electron density.
Thus, projecting interactions into this multiband manifold can stabilize a FCI state even at integer filling of the moir\'e unit cell.

 To demonstrate this mechanism, we consider one electron per moir\'e unit cell, $\nu=1$, and $Q=3$.
We refer to this setup as the three-flux model. 
Since the emergent flat-band manifold in Eq.~\eqref{wavefunction} contains three states per unit cell, this corresponds to an effective filling $1/3$ of that manifold, despite the integer filling of the unit cell. 

We therefore study interacting electrons projected to this three-band manifold, assuming that the single-particle gap to higher bands is sufficiently large to suppress interaction-induced band mixing. 
To expose the fractionalized liquid analytically, we consider the short-range repulsion 
$V(\r)= v_1 a^4\nabla^2\delta^{(2)}(\r)$ with $v_1=3V_1/(4\pi)$~\cite{Haldane1983,Trugman1985}, where FCI states emerge as zero energy modes~\cite{Jie_2021_LL,Ledwith_2020,Ledwith_2023}. 
Moreover, we parametrize the emergent magnetic field as $\mathcal B(\r)=\mathcal B_{n_{\Phi}}\left[1+\mathcal K\sum_{j=1}^3\cos(\G_j\cdot\r)\right]$ with $\G_{1,2,3}$ first shell of reciprocal lattice vectors and tunable $\mathcal K$.

Fig.~\ref{fig:bandsskyrmion}(b) displays the many-body spectrum obtained with ED on a torus at a finite modulation of the emergent magnetic field $\mathcal K=0.4$.   
The spectrum exhibits a threefold-degenerate manifold of \textit{exact} zero-energy ground states, separated from the excited states by a finite many-body gap, consistent with fractionalization and topological order.
For this short-range interaction, the zero modes can be constructed exactly~\cite{Jie_2021_LL,Ledwith_2020,Ledwith_2023} as generalized Laughlin states dressed by the skyrmion texture; their form is given in the SM~\cite{supplementary}.

This analytical zero-mode structure enables large-scale Monte Carlo simulations of strongly correlated states.
We characterize the topological properties by following the evolution of the many-body polarizations~\cite{Resta1998,Resta1999,Souza_2000}, 
$\zeta_j= \mel{\Psi_{\bm\phi}}{\exp\left(i\Delta\k_j\cdot\sum_{l=1}^N\hat \r_l\right)}{\Psi_{\bm\phi}}$ with $j=1,2$, under twisted boundary conditions on a torus with sides $\L_{1,2}$, momentum resolutions $\Delta\k_{1,2}$ and number of electrons $N$.  The twists are introduced by threading fluxes $\bm\phi=(\phi_1,\phi_2)$ through the two handles of the torus~\cite{supplementary}, which impose the boundary conditions $\Psi_{\bm\phi}(\r_1+\L_j,\cdots,\r_N)=e^{i\phi_j}\Psi_{\bm\phi}(\r_1,\cdots,\r_N)$. 
The evolution of the center-of-mass position along the direction $\L_{1,2}$, measured in units of the system size, is captured by the quantities $X_j = \Im\log \zeta_j/(2\pi)\in[-1/2,1/2)$.

In Fig.~\ref{fig:comflux_structurefactor}(a), we show $X_{1}$ and $X_2$ as a function of the threaded flux $\phi_2$. 
The Hall response is obtained from the charge pumped across the torus during the flux-insertion cycle from $0$ to $6\pi$, 
$C=h\sigma_{xy}/e^2=\int_0^{6\pi} d\phi_2\, \partial_{\phi_2}X_1/3$~\cite{Laughlin1981,Niu1985,chen2026topologicalinvariantperiodicbody}.
For a single branch of the ground-state manifold, insertion of one flux quantum pumps charge $1/3$ across the torus, yielding the Hall conductivity $\sigma_{xy}=e^2/(3h)$. 
This establishes a FCI state at the integer microscopic filling $\nu=1$, where the Hall response is no longer tied to the number of electrons per unit cell.

\textit{Perfect screening in the ideal FCI---} The generalized Laughlin wavefunction stabilized in our three-flux model retains the characteristic short-distance structure of the Laughlin state: when two electrons approach, the wavefunction develops a third-order zero, corresponding to three flux quanta effectively bound to each electron. 
This purely pairwise correlation structure allows the many-body probability distribution of ideal FCI to map exactly onto a classical two-dimensional Coulomb gas with \textit{two-body} logarithmic interactions~\cite{laughlin1983}. 
The inhomogeneous emergent magnetic field enters this plasma problem as an effective scalar potential~\cite{Wolf2025,Moitra2026,iyer2026dispersionanyonblochbands,yan2025anyondispersionaharonovcasherbands}.

This mapping directly relates the screening properties of the ground state to its long-wavelength density fluctuations, as captured by the connected static structure factor:
\begin{equation}
S(\q)=(\bra{\Psi} \hat\rho(\q)\hat\rho(-\q)\ket{\Psi} -|\bra{\Psi}\hat \rho(\q)\ket{\Psi}|^2)/A,    
\end{equation}
where $A$ is the area of the torus, and $\hat \rho(\q)=\sum_{j=1}^{N}e^{-i\q\cdot\hat\r_j}$ is the \textit{full} density operator. 
For incompressible phases, in the long-wavelength limit, the static structure factor vanishes quadratically, $S(\q)=\kappa\q^2/(4\pi)+\cdots$, where the coefficient~\cite{Onishi2025} $\kappa$ is bounded from below by the many-body Chern number $\kappa\ge |C|$~\cite{Onishi2024,ghosh2024probingquantumgeometryoptical}.

The ideal FCI corresponds to the screening phase of the classical Coulomb plasma. 
In this phase, density correlations are short ranged and decay exponentially in real space. The dielectric response of the plasma is encoded in the long-wavelength structure factor through
$\epsilon^{-1}=1-\frac{4\pi}{|C|}\lim_{\q\to 0}S(\q)/|\q|^2$~\cite{Martin1988,Moitra2026}. 
Perfect screening corresponds to $\epsilon^{-1}=0$, consistent with the exponential screening of density fluctuations. 
We therefore conclude directly from the plasma mapping that the ideal FCI saturates the bound, $\kappa=|C|$. 
$\kappa$ being proportional to the quantum metric~\cite{Onishi2025}, this saturation is the many-body analogue of the trace condition derived for lowest Landau levels and ideal bands~\cite{roy2014,Jackson_2015,Ledwith_2020,Jie_2021_LL,Ledwith_2023}.

Such saturation is not generic to fractional Chern insulators~\cite{Zaklama2025,guerci2026topologicalsuperconductivityemergentvortex}; it follows here from the exact Laughlin-type zero-mode structure and the associated perfect-screening plasma. 
This prediction is verified in Fig.~\ref{fig:comflux_structurefactor}(b) from the long-wavelength behavior of the static structure factor.  
Before concluding, we emphasize that the perfect-screening condition is also satisfied by ideal FCIs with a single flux quantum per unit cell, $Q=1$; see the SM~\cite{supplementary}.

\begin{figure}
    \centering
    \includegraphics[width=\linewidth]{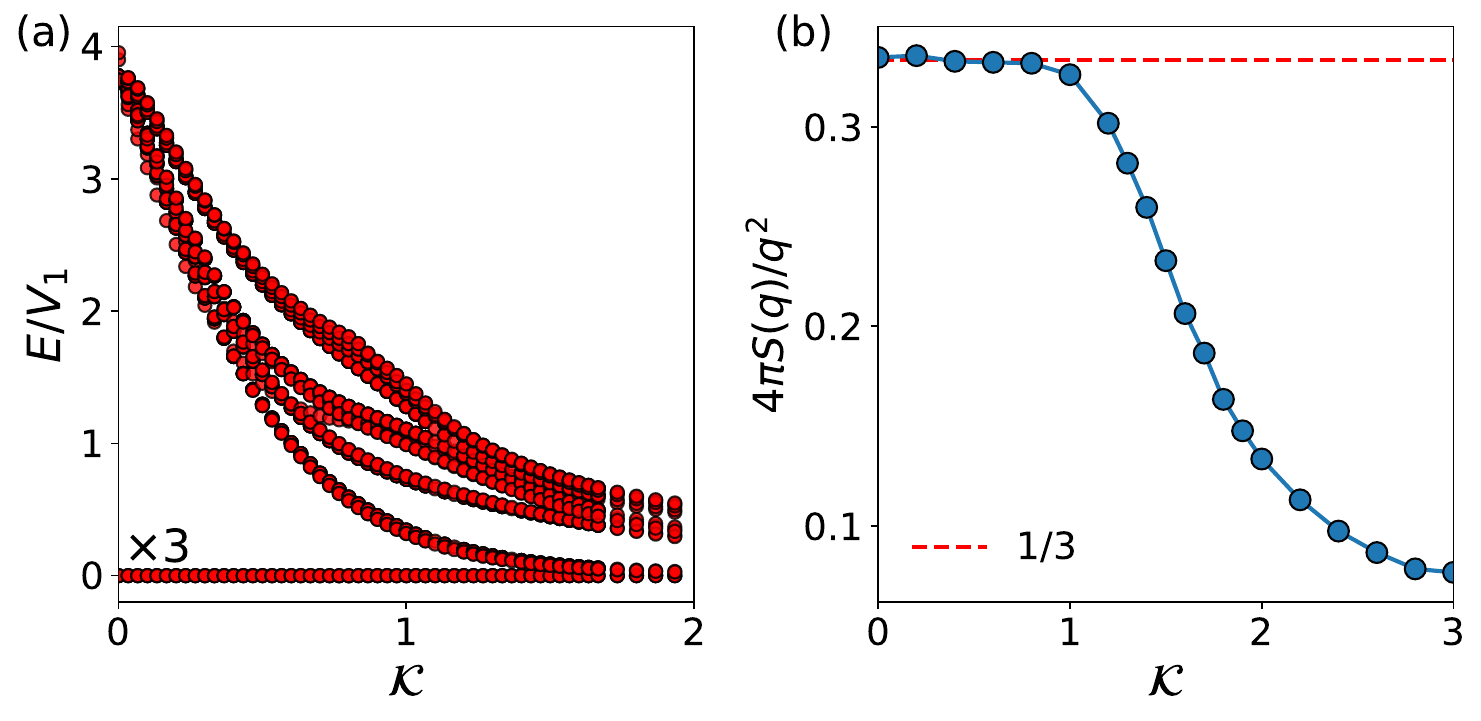}
\caption{\textbf{Evolution of the many-body spectrum and of the long wavelength behavior of $S(\q)$.} (a) Spectrum versus magnetic-field modulation $\mathcal K$ for $V_1$ interactions at $\nu=1$ and emergent flux $3h/e$ per unit cell. The calculation retains the three lowest bands on a $3\times3$ cluster. (b) $4\pi S(\q)/\q^2$ at $\q=\Delta\k$, with $\Delta\k$ momentum resolution, as a function of $\mathcal K$ for $N=243$ electrons.}    \label{fig:manybodygap}
\end{figure}

\begin{figure*}[t]
    \centering
    \includegraphics[width=\linewidth]{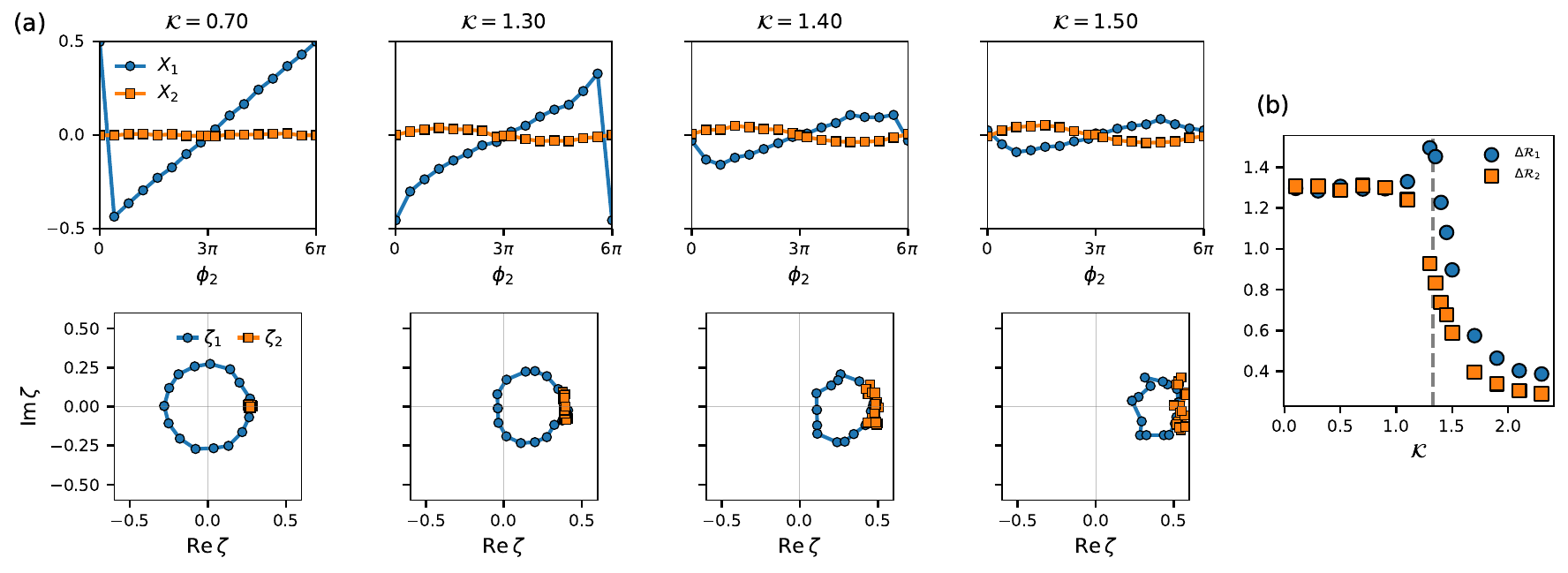}
    \caption{\textbf{Many-body polarization and center of mass fluctuations}. Panel (a): Top and bottom panels show the evolution of $X_{1/2}$ and $\zeta_{1,2}$, respectively, as a function of $\phi_2\in[0,6\pi)$ upon increasing $\mathcal K$. 
    Panel (b): Evolution of the twist averaged center-of-mass fluctuations $\Delta\mathcal{R}_{1,2}$. The vertical dashed line denotes the topological transition. Plots are for $N=243$ electrons.}
    \label{fig:many_body_winding}
\end{figure*}

\textit{Localization of the FCI ---} A natural question then concerns the fate of the FCI liquid as the emergent magnetic field becomes increasingly inhomogeneous.
Physically, the factor $e^{- K(\r)}$ in Eq.~\eqref{wavefunction} progressively concentrates the single-particle wavefunction near the minima of $K(\r)$. 
This enhanced localization promotes charge ordering and may ultimately destabilize the FCI liquid in favor of a dielectric state \cite{Clerouin87, Alastuey88}.
In particular, when each localization region encloses three flux quanta, $Q=3$, the $\mathcal K\to\infty$ limit corresponds to a crystal of localized electrons pinned at the minima of $K(\r)$.

We now provide analytical guidance showing that the competition between the crystalline dielectric and the FCI liquid is governed by BKT physics.
At large but finite $\mathcal K$, fluctuations about the pinned-crystal configuration occur through the formation of dilute electron-hole dipoles: an electron leaves its lattice site and moves close to a neighboring occupied site, leaving behind a vacancy. 
The resulting dilute dipole gas is described by the neutral Coulomb-gas model \cite{supplementary}, whose long-distance behavior is governed by the renormalization-group equations
\begin{equation}
\frac{d\epsilon^{-1}}{dl}
=
-12\pi^2 \epsilon^{-2} y^2,
\qquad
\frac{dy}{dl}
=
\left(2- \frac{3}{\epsilon}\right)y .
\end{equation}
The cost of creating a dipole grows with $\mathcal K$, causing the fugacity $y$ to be exponentially suppressed and to vanish as $\mathcal K\to\infty$. 
Thus, dipole excitations are absent in the pinned-crystal limit and remain confined at large but finite $\mathcal K$, corresponding to a dielectric phase with $\epsilon^{-1}=1$. 
Reducing $\mathcal K$ makes dipole formation favorable and renormalizes $\epsilon^{-1}$, driving the transition to a perfectly screening plasma with $\epsilon^{-1}=0$. 
At the transition, $\epsilon^{-1}$ exhibits the universal Nelson–Kosterlitz jump~\cite{Nelson1977}; see also Ref.~\cite{drouintouchette2022kosterlitzthoulessphasetransitionintroduction} for a recent review. Specifically, $\epsilon^{-1}$ jumps from $\epsilon^{-1}=2/3$ on the dielectric side to $\epsilon^{-1}=0$ in the plasma phase, in agreement with Ref.~\cite{Moitra2026}.
As shown in the SM~\cite{supplementary}, the detailed form of $K(\r)$ controls the bare defect fugacity, and hence the transition point, but not the long-distance physics.
This behavior sharply contrasts with the $Q=1$ case at $\nu=1/3$, where the FCI remains perfectly screening even at strong $\mathcal K$, rather than evolving into a pinned crystal.  
This phase is described by an exact many-body ground state wavefunction where electrons move on a lattice of three sites per electron; see Ref.~\cite{ji2026generalizedmodelfractionalquantum} and the SM~\cite{supplementary} for details. 

Guided by these analytical insights, we turn to numerical calculations and first examine the evolution of the many-body spectrum obtained from ED of finite-size systems.
Fig.~\ref{fig:manybodygap}(a) shows a pronounced softening of the many-body gap as $\mathcal K$ increases, while the gap remains finite on accessible system sizes. The three FCI zero modes remain at $\Gamma$, while eight low-energy states, one from each of the other many-body momentum sectors, progressively approach the ground-state manifold.

Beyond finite-size ED, we perform large-scale Monte Carlo simulations of $4\pi S(\q)/\q^2$ at $\q=\Delta\k$, with $\Delta\k$ momentum resolution of the torus. As shown in Fig.~\ref{fig:manybodygap}(b), this quantity is pinned to $1/3$ for $\mathcal K\lesssim1$, reflecting the perfect screening of the FCI plasma. 
Upon increasing $\mathcal K$, it decreases continuously toward zero, signaling the loss of screening and the approach to the dielectric limit in which electrons become localized around the minima of $\mathcal B(\r)$. 
The same evolution is observed across different system sizes~\cite{supplementary}, providing strong numerical support for the plasma-to-dielectric transition predicted by the Coulomb-gas analysis, although finite-size effects prevent a precise determination of the BKT transition point.

Remarkably, the plasma-to-dielectric transition is accompanied by a change in many-body topology. 
As shown in Fig.~\ref{fig:many_body_winding}(a), for representative values of $\mathcal K$ $(\mathcal K=0.7,\,1.3)$ the trajectory of $\zeta_1$ encircles the origin under flux insertion, yielding $\sigma_{xy}=e^2/(3h)$. On the other hand, for larger $\mathcal K$ $(\mathcal K=1.4,\,1.5)$, the trajectory no longer encloses the origin, signaling the loss of fractional charge pumping and the onset of a topologically trivial dielectric. 
Moreover, for a given system size, $|\zeta_j|\to1$ as $\mathcal K$ increases, consistent with increasing electron localization~\cite{Resta1999,Zhang2020}.
The change in topological response thus provides a sharp finite-size marker of the transition.

A complementary signature emerges from the flux-averaged fluctuations of the many-body center of mass,
$\Delta\mathcal{R}_j=-\int_0^{6\pi}d\phi_2\,\log|\zeta_j|/(6\pi)$, shown in Fig.~\ref{fig:many_body_winding}(b).
These fluctuations are large in the topological phase and drop sharply upon entering the dielectric regime.
Most notably, $\Delta\mathcal{R}_1$ develops a pronounced peak as the polarization trajectory approaches the origin (see $\mathcal K=1.3$ in Fig.~\ref{fig:many_body_winding}(b)), where $\log|\zeta_1|$ becomes singular.
Together, these results link the winding of the many-body polarization to the dielectric response of the BKT transition. The corresponding center-of-mass fluctuations constitute an indicator of the transition that develops a pronounced peak when the polarization trajectory crosses the origin, with a finite-size-dependent location in $\mathcal K$. 

\textit{Discussion ---} We have demonstrated that higher-winding magnetic textures ($Q > 1$) offer a route for engineering fractionalized topological phases at integer filling ($\nu=1$) of the microscopic unit cell. 
By decoupling microscopic filling from the emergent flux density, this mechanism provides a natural explanation for incompressible phases whose filling is not uniquely determined by their Hall response, as encoded by the Streda relation in the presence of an external magnetic field. Experimental observation of fractional phases where the filling is not tied to the many-body Chern number has already been reported \cite{Li2026}. 

Increasing the inhomogeneity of the emergent magnetic field drives the higher-winding FCI into a dielectric phase adiabatically connected to a topologically trivial crystal of pinned electrons. 
This phase is naturally understood in the strong-modulation limit, where three flux quanta localize near each site and bind one electron to each local flux configuration.
In the Coulomb-gas description, this transition corresponds to the confinement or proliferation of dipole defects and is governed by BKT physics.

The same transition is visible directly in the many-body polarization: the quantized winding under flux insertion is lost upon entering the dielectric phase, while the associated center-of-mass fluctuations peak near the transition. 
These observables provide practical finite-size diagnostics of the transition.

Higher-winding textures could potentially be realized in engineered optical flux lattices or multi-layer moiré heterostructures~\cite{kaplan2025machinelearningassistedhigh,nakatsuji2025highthroughputdiscoverymoirehomobilayers,xu2026organizingprinciplesmoirequantum}. 
A key next step is to identify realistic platforms in which fractional states can emerge at integer unit-cell filling, enabling access to the plasma-to-dielectric transition discussed here. 
Quantifying the robustness of the integer-filling FCI and its BKT transition against finite band dispersion and longer-range interactions remains an important open direction.

\textit{Acknowledgements---} We thank Ahmed Abouelkomsan, Filippo Gaggioli, Liang Fu, Yugo Onishi, Nicolas Regnault, and especially Inti Sodemann, whose insights stimulated our interest in this problem. 
We acknowledge MIT SuperCloud and the Lincoln Laboratory Supercomputing Center for providing computational resources.

\bibliography{biblio}


\onecolumngrid
\newpage
\makeatletter 

\begin{center}
\textbf{\large Supplementary Materials for: ``\@title ''} \\[10pt]
Kishore Iyer,$^1$ Christophe Mora,$^1$ and Daniele Guerci$^{2,3}$ \\
\textit{$^1$Universit\'e Paris Cit\'e, CNRS,  Laboratoire  Mat\'eriaux  et  Ph\'enom\`enes  Quantiques, 75013  Paris,  France}\\
\textit{$^2$Department of Physics, William \& Mary, Williamsburg, VA-23187, USA}\\
\textit{$^3$Department of Physics, Massachusetts Institute of Technology, Cambridge, MA-02139, USA}\\
\end{center}
\vspace{10pt}

\setcounter{page}{1}  
\setcounter{figure}{0}
\setcounter{section}{0}
\setcounter{equation}{0}

\renewcommand{\thefigure}{S\@arabic\c@figure}
\makeatother

\appendix 

In these Supplementary Materials, we provide additional details on the theoretical results presented in the main text. 
Sec.~\ref{app_sec:single_particle} presents the higher-winding skyrmion-texture Hamiltonian, with examples for square and triangular lattice geometries. 
In Sec.~\ref{app_sec:many_body}, we introduce the many-body Hamiltonian used in the exact diagonalization calculations. 
Sec.~\ref{app_sec:MC_Laughlin} provides details on our Monte Carlo simulations and the generalized Laughlin wavefunction on the torus for the higher flux system. Sec. ~\ref{app:one_flux} provides results on the one-flux system.
Finally, in Sec.~\ref{app_sec:neutral_gas}, we derive the BKT renormalization group equations starting from the many-body wavefunction.

\section{Higher-Winding Skyrmion Texture Hamiltonian}
\label{app_sec:single_particle}

We consider two-component electrons subject to a skyrmion texture:
\begin{equation}\label{app:skyrmionmodel}
H(\r)=-\frac{\hbar^2\nabla^2}{2m_*}+J\bm n(\r)\cdot\bm\sigma+V(\r),
\end{equation}
where $\bm n(\r)$ is a unit vector describing the local skyrmion texture, $J$ is the exchange coupling, $\bm\sigma$ denotes the Pauli matrices acting on the internal degree of freedom, and $V(\r)$ is a lattice-periodic scalar potential.
We measure energies in units of the kinetic scale $\hbar^2/(m_*a^2)$, lengths in units of $a$, and wave vectors in units of $a^{-1}$, with $a$ the unit cell side length. We note that throughout this Appendix we focus on $\mathrm{SU}(2)$ Hamiltonians. Their generalization to $N>2$ internal degrees of freedom can be carried out by replacing the Pauli matrices with the generators of $\mathrm{SU}(N)$ and taking into account their corresponding algebra~\cite{Graf2021}.

In the large-exchange limit, we project onto the local pseudospin state $\vec\chi$ with exchange energy $-J$. 
The eigenstate is represented by spin-SU(2) coherent states:
\begin{equation}
\vec \chi = \sqrt{\frac{1}{2-2n_z}}(1-\bm n\cdot\bm\sigma)\begin{pmatrix}
    1 \\
    0
\end{pmatrix}.    
\end{equation}
After subtracting this constant exchange energy, the resulting dimensionless scalar Hamiltonian reads
\begin{equation}
\mathcal H(\r)=\frac{m_*a^2}{\hbar^2}\left[\vec\chi^\dagger(\r) H(\r) \vec\chi(\r) + J\right]=
\frac{\left[-i\partial_j+\mathcal A_j(\r)\right]
\left[-i\partial_j+\mathcal A_j(\r)\right]}{2}
+D(\r)
+\frac{m_* a^2}{\hbar^2}V(\r),
\end{equation}
where the energy is measured in units of $\hbar^2/(m_* a^2)$ and we have introduced the emergent vector potential: 
\begin{equation}
\mathcal A_j=-i\vec\chi^\dagger\partial_j\vec\chi=(\bm n\times\partial_j\bm n)_z/(2-2n_z).     
\end{equation}
In addition, we have introduced $D(\r)$ which can be interpreted as the trace of the real-space quantum metric:
\begin{equation}
    D(\r)=\frac{1}{2}\sum_j\left[\left(\partial_j\vec\chi^\dagger\right)(\partial_j\vec\chi)-(\partial_j\vec\chi^\dagger) \cdot\vec\chi\vec\chi^\dagger\cdot(\partial_j\vec\chi)\right]=\frac{1}{8}\sum_j\partial_j\bm n(\r)\cdot \partial_j\bm n(\r).
\end{equation}
In addition, we introduce the emergent magnetic field: 
\begin{equation}
    \mathcal B(\r) = \partial_x\mathcal  A_y(\r)-\partial_y\mathcal A_x(\r)=-\frac{1}{2}\bm n(\r)\cdot[\partial_x\bm n(\r)\times\partial_y\bm n(\r)].
\end{equation}
In the following, we assume the magnetic flux per unit cell is negative: 
\begin{equation}
    Q=\frac{1}{4\pi}\int_{\rm UC}d^2\r ~\bm n(\r)\cdot[\partial_x\bm n(\r)\times\partial_y\bm n(\r)]=-n_{\Phi}>0.
\end{equation}
In this case, the zero mode Dirac operator reads $\mathcal D=-i\bar \partial+\mathcal A$, which annihilates the holomorphic sector. The resulting Hamiltonian can be expressed as: 
\begin{equation}
\mathcal H(\r)=
\frac{\mathcal D^\dagger\mathcal D}{2}
-\frac{\mathcal B(\r)}{2}
+D(\r)
+\frac{m_* a^2}{\hbar^2}V(\r).
\end{equation}
Finally, the Aharonov--Casher limit is obtained by setting the potential to: 
\begin{equation}
    V(\r)=\frac{\hbar^2\mathcal B(\r)}{2m_* a^2}-\frac{\hbar^2 D(\r)}{m_* a^2}=-\frac{\hbar^2}{8m_* a^2}\left[2\bm n(\r)\cdot[\partial_x\bm n(\r)\times\partial_y\bm n(\r)]+ \sum_j\partial_j\bm n(\r)\cdot \partial_j\bm n(\r)\right].
\end{equation}
The latter choice of $V(\r)$ leads, in the $J\to\infty$ limit, to the Aharonov--Casher Hamiltonian:
\begin{equation}\label{app:ahronovcasher}
\mathcal H_{\rm AC}(\r)=
\frac{(-i \partial+\mathcal A(\r)^*)(-i\bar \partial+\mathcal A(\r))}{2}.
\end{equation}

We conclude by observing that, for a positive magnetic flux, the zero mode belongs to the antiholomorphic sector and satisfies $(-i \partial+\mathcal A(\r)^*)\psi=0$.

\subsection{Square geometry}

We consider a layer skyrmion texture with square lattice periodicity $\G_1=(2\pi/a,0)$ and $\G_2=(0,2\pi/a)$ and components:
\begin{equation}\label{app:square_exchange}
    (J_x+iJ_y)/J=(\sin\G_1\cdot\r+i\sin\G_2\cdot\r)^3,\quad J_z/J=m+\cos(\G_1\cdot\r)+\cos(\G_2\cdot\r),\quad 
\bm n (\r) = \frac{\bm J(\r)}{|\bm J(\r)|},
\end{equation}
with $J$ exchange coupling constant and $m$ measured in units of $J$.
The corresponding Pontryagin index is $Q=3$ when $-2<m<0$ corresponds to a negative net magnetic flux per unit cell. 

\begin{figure}
    \centering
    \includegraphics[width=\linewidth]{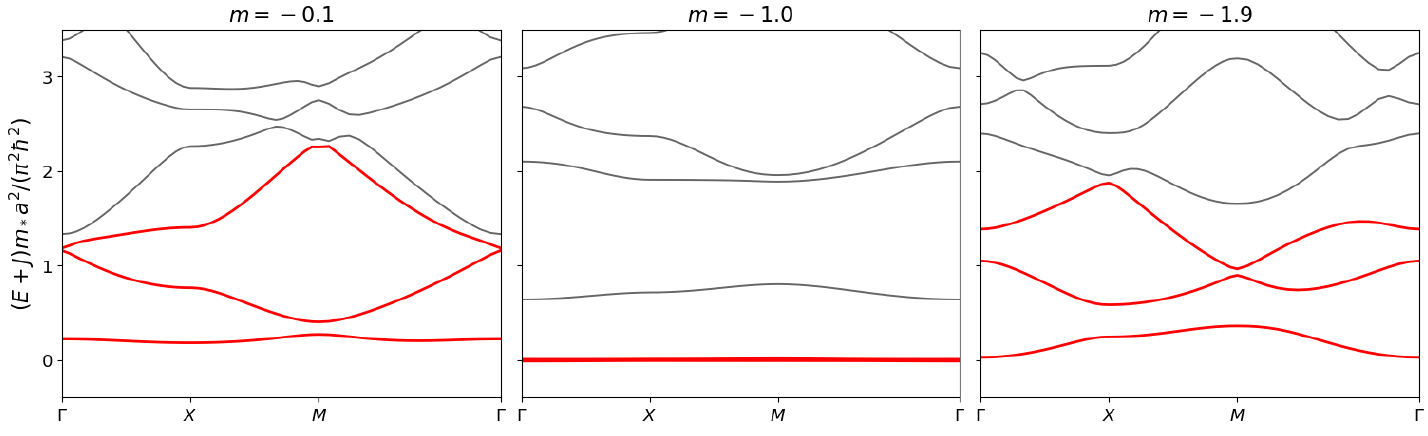}
    \caption{Energy spectrum of the square lattice higher-winding skyrmion model in the strong-exchange coupling $J$ regime, $J m_* a^2/(\pi^2\hbar)^2=500$.}
    \label{energy_dispersion_square}
\end{figure}

\begin{figure}
    \centering
    \includegraphics[width=.8\linewidth]{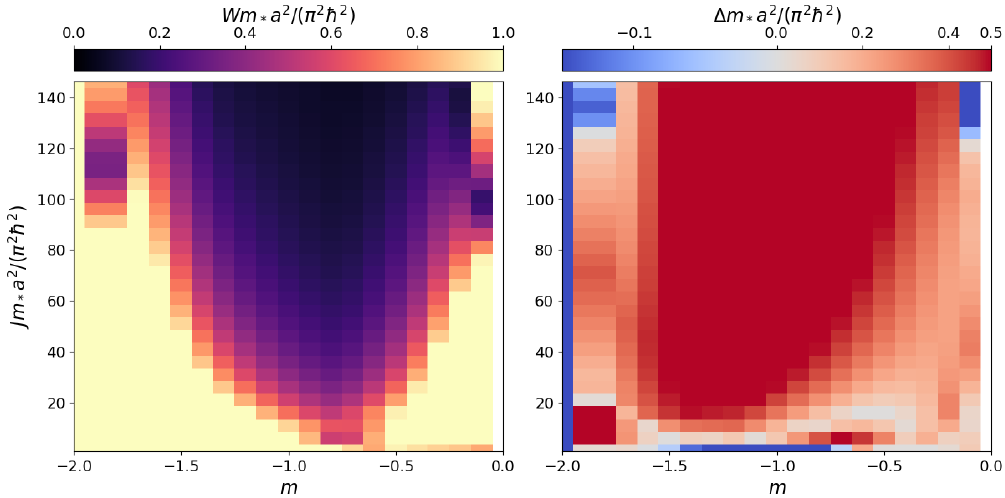}
    \caption{Phase diagram of the higher-winding skyrmion model as a function of $m$ and $J$ for a square lattice, characterized through the dispersion and isolation of the three lowest-energy bands. The left panel shows their bandwidth $W$, while the right panel shows the energy gap $\Delta$ separating them from the remote bands.}    
\label{fig:phase_diagram_square}
\end{figure}

Fig.~\ref{energy_dispersion_square} shows the band structure for $J m_* a^2/(\pi^2\hbar)^2=500$ and three different values of the ``mass term" $m$ in Eq.~\eqref{app:square_exchange}. 
At the center of the topological region, for $m=-1$, the spectrum contains a $Q$-fold degenerate manifold of flat bands. 
On the other hand, as we approach the lower $m=-1.9$ or the upper boundary $m=-0.1$, the bands become increasingly dispersive, and the three low-energy bands are no longer isolated from the higher-energy states.
Fig.~\ref{fig:phase_diagram_square} characterizes the evolution of the band dispersion as a function of $m$ and $J$. 
To this aim, we introduce the bandwidth of the $Q=3$ lowest energy bands: 
\begin{equation}\label{app:bandwidth}
    W=\max_{\k\in \rm BZ} \epsilon_{3}(\k)-\min_{\k\in \rm BZ} \epsilon_{1}(\k),
\end{equation}
and the gap from the remote bands: 
\begin{equation}\label{app:gap}
    \Delta=\min_{\k\in\rm BZ} \epsilon_{4}(\k)-\max_{\k\in\rm BZ} \epsilon_{3}(\k).
\end{equation}
The left panel of Fig.~\ref{fig:phase_diagram_square} shows the bandwidth $W$ of the three lowest-energy bands, while the right panel of Fig.~\ref{fig:phase_diagram_square} shows their gap $\Delta$ to remote bands.

\subsection{Triangular geometry}

\begin{figure}
    \centering
    \includegraphics[width=\linewidth]{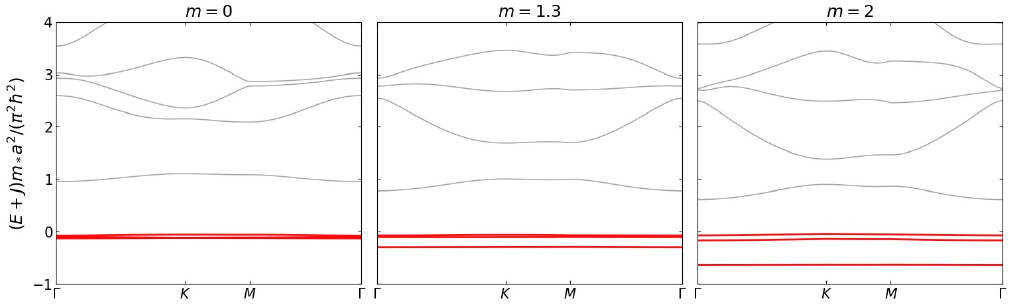}
    \caption{Energy spectrum of the triangular lattice higher-winding skyrmion model in the strong-exchange coupling $J$ regime, $J m_* a^2/(\pi^2\hbar)^2=100$. }
    \label{energy_dispersion_triangular}
\end{figure}

\begin{figure}
    \centering
    \includegraphics[width=.8\linewidth]{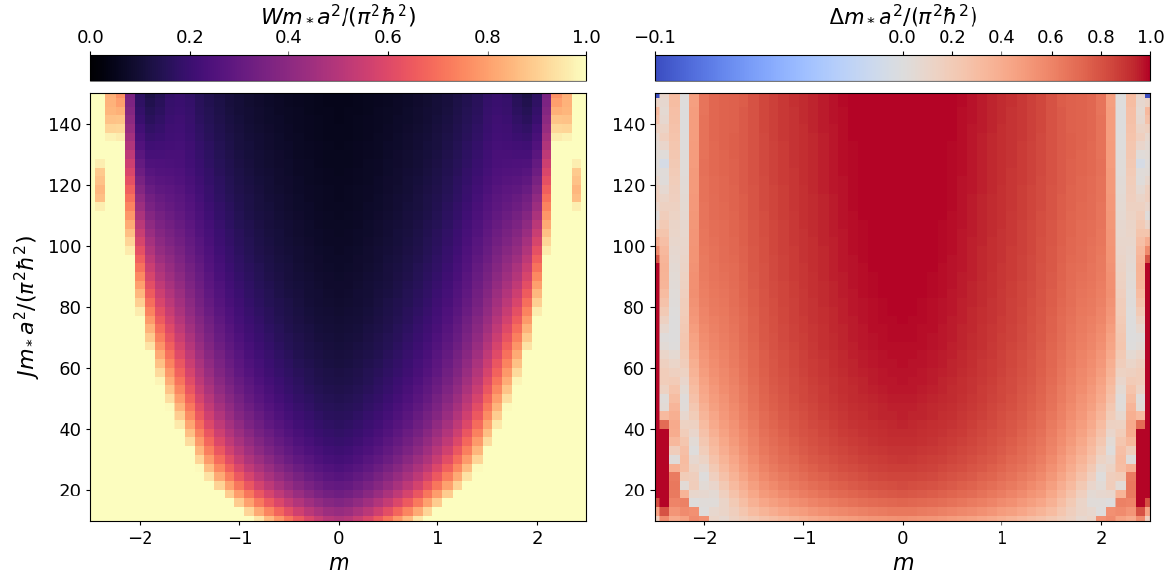}
    \caption{Phase diagram of the higher-winding skyrmion model as a function of $m$ and $J$ for a triangular lattice, characterized through the bandwidth and isolation of the three lowest-energy bands. The left panel shows their bandwidth $W$, while the right panel shows the energy gap $\Delta$ separating them from the remote bands.}    
\label{fig:phase_diagram_triangular}
\end{figure}

We consider a layer skyrmion texture with triangular lattice periodicity, with Dirac wave vectors $\bm K_j=4\pi/(3a)\left(-\sin2\pi j/3,\cos2\pi j/3\right)$ in complex notation and reciprocal lattice vectors $\G_j=\bm K_j-\bm K_{j+1}$. 
The components of the Zeeman pseudospin field are as follows:
\begin{equation}\label{app:triangular_exchange}
    (J_x+iJ_y)/J=\left(\sum_{j=1}^3 e^{i\bm K_j\cdot\r}\right)^3,\quad J_z/J=m-\sum_{j=1}^{3}\sin(\G_j\cdot\r),\quad 
\bm n (\r) = \frac{\bm J(\r)}{|\bm J(\r)|},
\end{equation}
with $J$ exchange coupling constant and $m$ measured in units of $J$.
The corresponding Pontryagin index is $Q=3$ when $-3\sqrt{3}/2<m<3\sqrt{3}/2$ corresponds to a negative net magnetic flux per unit cell. 
Fig.~\ref{energy_dispersion_triangular} shows the spectrum for several values of $m$ at $J m_* a^2/(\pi^2\hbar)^2=100$.
Similarly to the square-lattice model, the most robust flat bands occur near the center of the topological domain, where they remain weakly dispersive and well separated from the remote bands.
Fig.~\ref{fig:phase_diagram_triangular} shows the single-particle phase diagram of the triangular-lattice model. The left panel displays the bandwidth $W$~\eqref{app:bandwidth} of the lowest-energy bands, while the right panel shows their energy gap $\Delta$~\eqref{app:gap} from the remote bands.

We observe that for $|m|>3\sqrt{3}/2$, the net flux through the unit cell vanishes. As a consequence, the $Q$-fold quasi-degenerate flat-band manifold is no longer realized, even in the large-$J$ limit. 

\section{Many-body Hamiltonian in the Aharonov-Casher limit}
\label{app_sec:many_body}

We consider interacting electrons described by the Hamiltonian:   
\begin{equation}
    H(\r_1,\cdots,\r_N)=\sum_{j=1}^N \frac{\hbar^2[-i \partial_j+\mathcal A(\r_j)^*][-i\bar \partial_j+\mathcal A(\r_j)]}{2m_*} + \sum_{i<j} V(\r_i-\r_j),
\end{equation}
where $V(\r)$ is a two-body interaction.  
In second quantization, the Hamiltonian projected onto the $Q$-fold degenerate bands takes the form:
\begin{equation}
\begin{split}
    \hat H
    =\frac{1}{2A}\sum_{n_1\cdots n_4=1}^{Q}\sum_{\k_1\cdots \k_4}H^{n_1n_2,n_3n_4}_{\k_1\k_2,\k_3\k_4}\, c^\dagger_{\k_1 n_1}c^\dagger_{\k_2n_2}c_{\k_3n_3}c_{\k_4n_4},
\end{split}
\end{equation}
where we have introduced $A=|\L_1\times \L_2|$ the area of the torus and the interaction matrix element: 
\begin{equation}\label{matrix_element}
    H^{n_1n_2,n_3n_4}_{\k_1\k_2,\k_3\k_4}= \sum_\g \delta_{\k_1+\k_2-\k_3-\k_4,\Delta \g}\, V_{\k_1-\k_4-\g} \,\Lambda^{n_1,n_4}_{\k_1,\k_4+\g}\,\Lambda^{n_2,n_3}_{\k_2,\k_3+\Delta \g-\g}.
\end{equation}
  The index $n_j=1,\ldots,n_{\Phi}$ labels the $n_{\Phi}=|Q|\in\mathbb N$ degenerate bands. In the previous expression, we have introduced the form factor: 
\begin{equation}\begin{split}
    \Lambda^{n,n'}_{\k,\p+\G} = \int_{\rm UC}\frac{d^2\r}{|\a_1\times\a_2|}e^{i(\k-\p-\G)\cdot\r}\vec \psi^\dagger_{\k n}(\r)\cdot\vec\psi_{\p n'}(\r)=\left[\mathcal N^\dagger(\k)\sum_{\G'}w_{\G'}\left(f^{\k,\p}_{\G+\G'}\right)\mathcal N(\p)\right]_{n,n'},
\end{split}\end{equation}
where we have introduced the matrix: 
\begin{equation}
    \left[f^{\k,\p}_{\G}\right]_{n,n'}=\int_{\rm UC}\frac{d^2\r}{|\a_1\times\a_2|}\Phi^*_{\k n}(\r) e^{i(\k-\p-\G)\cdot \r}\Phi_{\p n'}(\r)=\mel{\k,n}{e^{i(\k-\p-\G)\cdot\r}}{\p,n'},
\end{equation}
and we have defined the Fourier coefficients:
\begin{equation}
    w_{\G}=\int_{\rm UC}\frac{d^2\r}{|\a_1\times\a_2|}e^{i\G\cdot\r}e^{-2K(\r)}.
\end{equation}
The normalization matrix $\mathcal N(\k)$ is obtained imposing the condition: 
\begin{equation}
    \Lambda^{n,n'}_{\k,\k}=\int_{\rm UC}\frac{d^2\r}{|\a_1\times\a_2|}\vec \psi^\dagger_{\k n}(\r)\vec \psi_{\k n'}(\r)=\delta_{nn'}.
\end{equation}
We find: 
\begin{equation}
    \mathcal N^\dagger (\k) S(\k) \mathcal N(\k)=1,\quad \mathcal N(\k)=S^{-1/2}(\k),
\end{equation}
where we have introduced the overlap: 
\begin{equation}\begin{split}
    S_{nn'}(\k)&=\int_{\rm UC} \frac{d^2\r}{|\a_1\times\a_2|}e^{-2K(\r)}\Phi^*_{\k n}(\r)\Phi_{\k n'}(\r) =\sum_{\G}w_{\G}\left( f^{\k,\k}_{\G}\right)_{nn'}.
\end{split}\end{equation}

The evaluation of this matrix element is performed by exploiting magnetic translation algebra. 
The final result reads:
\begin{equation}
    \left[f^{\k,\p}_{\G}\right]_{n,n'}=e^{i\ell^2(\p+\k)\times\G/(2n_{\Phi})}e^{i\ell^2\k\times\p/(2n_{\Phi})}e^{-|\k-\p-\G|^2\ell^2/(4n_\Phi)}\left[\eta^{(n_{\Phi})}_{\G}\right]_{n,n'},
\end{equation}
where $\eta^{(n_{\Phi})}_{\G}$ is the signature.  
Given $\G=n_1\b_1+n_2\b_2$, for $n_{\Phi}=1$, we find: 
\begin{equation}
    \eta^{(1)}_{\G}=e^{-i(n_1+n_2+n_1n_2)\pi}.
\end{equation}
On the other hand, for $n_{\Phi}>1$, we have: 
\begin{equation}
\left[\eta^{(n_{\Phi})}_{\G}\right]_{n,n'}=\left[\left(\sigma\right)^{-n_1}\tau^{n_2}\right]_{n,n'}e^{-in_1n_2\frac{\pi}{n_\Phi}}e^{i\frac{\pi}{3n_{\Phi}}(n_2-n_1)\,\text{mod}(n_{\Phi},2)},
\end{equation}
where we have introduced the clock and shift matrices $\tau$ and $\sigma$~\cite{GuerciPRB2025}: 
\begin{equation}
    \sigma^{-1} = \begin{pmatrix}
        0 & 1 & \ldots  & 0 \\
        \vdots  & \ddots &  \ddots & \vdots \\
        0 &   \ldots & \ldots & 1  \\
        1 &  0 & \ldots &  0
    \end{pmatrix},\quad \tau =\text{diag}[1,e^{i2\pi/n_\Phi},\cdots,e^{i2\pi(n_\Phi-1)/n_\Phi}]. 
\end{equation} 

Exact diagonalization simulations presented in the main text are performed for $n_\Phi=3$ by solving the many-body Hamiltonian at one electron per unit cell, retaining the lowest $n_\Phi$ bands. Specifically, we perform exact diagonalization in momentum space on a $3\times 3$ cluster with $3$ orbitals per $\k$ point and interaction $V(\r)= v_1 a^4\nabla^2\delta^{(2)}(\r)$ with $v_1=3V_1/(4\pi)$. 

\section{Monte Carlo simulations of the generalized Laughlin state}
\label{app_sec:MC_Laughlin}

To begin, we consider the generalized Laughlin wavefunction at filling $1/Q$ per flux quantum with twisted boundary conditions $\bm\phi=(\phi_1,\phi_2)$. 
The many-body wavefunction reads: 
\begin{equation}\label{psi_gstate}\begin{split}
    \Phi_{\bm\phi}(\x_1,\cdots,\x_N) &= e^{i\R\times\bm(\phi_1\L_2-\phi_2\L_1) /(4\pi Q N\ell^2_B)} \prod_{i<j} f^{Q}(\r_i-\r_j)\prod_{l=1}^{Q} f\left(\R+\frac{1}{2\pi}\frac{\phi_1\L_2-\phi_2\L_1}{Q}\right)\prod_{i=1}^{N}e^{-K(\r_i)}\chi_{\ell_j}(\r_i)\\
    &=\Psi_{\bm\phi}(\r_1,\cdots,\r_N) \prod_{i=1}^{N}\chi_{\ell_j}(\r_i),
\end{split}\end{equation}
where $\x_j=(\r_j,\ell_j)$ denotes the combined coordinate, with $\r_j$ the position on the torus and $\ell_j$ the layer index, $\R=\sum_{j=1}^N \r_j$ is the center of mass coordinate, $N_s$ is the number of unit cells, $N/N_s=1/Q$, and $\ell_B^2=A_{\rm UC}/(2\pi)$ is the magnetic length associated with one flux quantum per unit cell.
Twisting the boundary is equivalent to applying the translation operator to the center of mass part of the wavefunction: 
\begin{equation}
    \hat T_{\X}=e^{i\X\cdot(\mathbf z \times \R)/(2\ell^2_B N)} e^{\X\cdot\nabla_\R},
\end{equation}
where $\nabla_\R=\sum_{j=1}^N \nabla_{j}$ acts on each particle coordinate. 

In Eq.~\eqref{psi_gstate}, the function $f(\r)$ satisfies the relation: 
\begin{equation}
    f(\r +\bm L)=\eta_{\L} e^{i\frac{\L\times \r}{2\ell^2_B N_s}}f(\r),
\end{equation}
where $\eta_{\L}=e^{-i\pi(n+m+nm)}$ is the signature of the lattice vector $\bm L=n\L_1+m\L_2$, with $\L_{1,2}$ supercell vectors. 
In our analysis, we consider supercells defined by $\L_{1,2}=\sqrt{N_s}\a_{1,2}$, with $N_s$ a perfect square, unit aspect ratio, and relative angle $2\pi/3$. Specifically, the function $f(\r)$ reads: 
\begin{equation}\label{f_wavefunction}
    f(\r) = \vartheta_1\left(\frac{z}{L_1},\omega\right)e^{\frac{\pi}{2\Im\omega}\left[\left(\frac{z}{L_1}\right)^2-\left|\frac{z}{L_1}\right|^2\right]},
\end{equation}
where $\omega=e^{2\pi i/3}$, $\r_j = (x_j, y_j), ~z_j = x_j + i y_j$, and we have introduced the Jacobi theta function: 
\begin{equation}    \vartheta_1(z,\omega)=\sum_{n\in\mathbb Z} e^{i \pi \omega (n+1/2)^2} e^{2 i \pi (z-1/2)(n+1/2)}.
\end{equation}
Upon advancing $\r_1\to\r_1+\L_j$ with $j=1,2$, we readily verify that the wavefunction satisfies the boundary conditions: 
\begin{equation}
    \frac{ \Psi_{\bm\phi}(\r_1+\L_j,\cdots,\r_N)}{ \Psi_{\bm\phi}(\r_1,\cdots,\r_N)}=\eta^{N_s}_{\L_j}e^{i\L_j\times\r_1/(2\ell^2_B)}e^{i\phi_j}.
\end{equation}

In our analysis, we consider periodic emergent magnetic-field modulations $\delta\mathcal B(\r)$ enclosing one and three flux quanta per unit cell.
For one flux quantum per unit cell, we considered the Kähler potential
\begin{equation}\label{one_flux}
K(\r) = -\frac{\sqrt{3}}{4\pi}\mathcal K \sum_{j=1}^{3}\cos(\G_j\cdot\r),
\end{equation}
where $\G_1=(4\pi/\sqrt 3a)$, $\G_2=(4\pi/\sqrt 3a)(-1/2,\sqrt{3}/2)$ and $\G_3=-\G_1-\G_2$. 
On the other hand, for three fluxes per unit cell, we have:
\begin{equation}\label{three_flux}
K(\r) = -\frac{3\sqrt{3}}{4\pi}\mathcal K \sum_{j=1}^{3}\cos(\q_j\cdot\r),
\end{equation}
where $|\q_1\times\q_2|/|\G_1\times\G_2|=1/3$ and the unit cell encloses three flux quanta.
We recall that the corresponding magnetic field is:
\begin{equation}\label{app:magnetic_field}
\mathcal B(\r) = \mathcal B_{n_{\Phi}}\left[1+\mathcal K\sum_{j=1}^3\cos\varphi_j(\r) \right].
\end{equation}
In the last expression, we have introduced the phase $\varphi(\r) = \k\cdot\r$ with $\k$ either $\G$ or $\q$.

Finally, before proceeding, we note that the center-of-mass part of Eq.~\eqref{psi_gstate} is characterized by a zero of order $Q$ located at $\R=(\L_1\phi_2-\L_2\phi_1)/(2\pi Q)$. The position of the zero moves as a function of the twist of the boundary conditions $(\phi_1,\phi_2)$.
There are high-symmetry configurations at specific boundary-condition twists where the zero is located at a high-symmetry position of the triangular supercell.  
These include $\bm\phi=(0,0)$, $\bm\phi=(0,Q\pi)$, $(Q\pi,0)$, and $(Q\pi,Q\pi)$. 
For the latter, the center of mass zero is located at $\R=(\L_1-\L_2)/2$, while the corresponding center of mass position lies at the origin, as it is conjugate to the zero. 
At this configuration, the center of mass exhibits the full point-group symmetry of the model. We therefore use it as our reference point to characterize the response of the ground state under flux threading.

\subsection{Evaluation of the charge density and density-density fluctuations}

 The average value of the single particle density reads: 
 \begin{equation}
     \langle \rho(\r) \rangle = \frac{N}{\braket{\Psi}{\Psi}}\int d^2\r_2\cdots d^2\r_N |\Psi(\r,\r_2,\cdots,\r_N)|^2.
 \end{equation}
As a result of the boundary conditions of the many-body wavefunction, we readily have $\langle\rho(\r+\a_j)\rangle=\langle\rho(\r)\rangle$ implying that the average density reads: 
\begin{equation}
    \langle \rho(\r) \rangle = \frac{1}{A}\sum_{\G} e^{i\G\cdot\r} \langle \rho_{\G}\rangle.
\end{equation}
To compute the density, we directly sample its Fourier components:
\begin{equation}
    \langle \rho_{\G}\rangle = \int d^{2}\{\x\} \sum_{j=1}^{N} e^{-i\G\cdot\r_j}P(\{\x\}),
\end{equation}
where $P=|\Psi|^2/\braket{\Psi}{\Psi}$ is the probability distribution. The expectation value is evaluated using Markov chain Monte Carlo,
\begin{equation}
    \langle \rho_{\G}\rangle \approx \mathbb E_{\Psi} \left[\sum_{j=1}^{N} e^{-i\G\cdot\r_j}\right]=\frac{1}{N_c}\sum_{n=1}^{N_c}\left[\sum_{j=1}^{N} e^{-i\G\cdot\r^{(n)}_j}\right].
\end{equation}

We define the connected density-density correlation function as: 
\begin{equation}
    S(\r,\r') = \langle \delta\rho(\r)\delta\rho(\r')\rangle,
\end{equation}
where $\delta\rho (\r) =\rho(\r) -\langle\rho(\r) \rangle$. 
Due to the boundary conditions, the connected density-density correlation function can be written as
\begin{equation}
    S(\r,\r') = \frac{1}{A^2}\sum_{\q\in \rm BZ}\sum_{\G\G'} e^{i\q\cdot(\r-\r')+i\G\cdot\r-i\G'\cdot\r'} \langle \delta \rho_{\q+\G}\delta \rho_{-\q-\G'}\rangle,
\end{equation}
where we have introduced $\delta\rho_{\q+\G}=\rho_{\q+\G}-\delta_{\q,0}\langle\rho_\G\rangle$. 
We introduce the quantity: 
\begin{equation}
    S_{\G,\G'}(\q) = \frac{\langle \delta \rho_{\q+\G}\delta \rho_{-\q-\G'}\rangle}{A},\quad \langle\rho_{\q+\G}\rho_{-\q-\G'}\rangle=\sum_{ij=1}^N\int d^2\{\x\}e^{-i\q\cdot(\r_i-\r_j)-i\G\cdot\r_i+i\G'\cdot\r_j}P(\{\x\}). 
\end{equation}
The previous average is computed statistically: 
\begin{equation}
    \langle\rho_{\q+\G}\rho_{-\q-\G'}\rangle\approx \mathbb E_{\Psi} \left[\sum_{ij=1}^{N}e^{-i(\q+\G)\cdot\r_i}e^{i(\q+\G')\cdot\r_j}\right]=\frac{1}{N_c}\sum_{n=1}^{N_c} \left(\sum_{i=1}^N e^{-i(\q+\G)\cdot\r^{(n)}_i}\right)\left(\sum_{j=1}^N e^{i(\q+\G')\cdot\r^{(n)}_j}\right).
\end{equation}

To study the long-wavelength response, we introduce the cell-average $S_{\rm av}(\r)$ defined as: 
\begin{equation}
    S_{\rm av}(\r)= \int_{\rm UC} \frac{d^2\x}{A_{\rm UC}} S(\r+\x,\x)=\frac{1}{A}\sum_{\q\in\rm BZ}\sum_{\G}e^{i(\q+\G)\cdot\r}S_{\G,\G}(\q)=\frac{1}{A}\sum_{\q}e^{i\q\cdot\r}S(\q),
\end{equation}
where in the last step $\q$ is unconstrained:
\begin{equation}
    S(\q) = \frac{\langle\rho_\q\rho_{-\q}\rangle-\delta_{\q,\G}|\langle\rho_{\G}\rangle|^2}{A}.
\end{equation}
The dielectric constant is obtained as: 
\begin{equation}
    \epsilon^{-1}=1-\lim_{\q\to0} \frac{4\pi Q S(\q)}{|\q|^2}. 
\end{equation}
In the plasma regime, $\epsilon^{-1}=0$ corresponds to perfect screening, while it becomes finite in the dielectric phase.

\subsection{Pair correlation function}

\begin{figure}
    \centering
    \includegraphics[width=.8\linewidth]{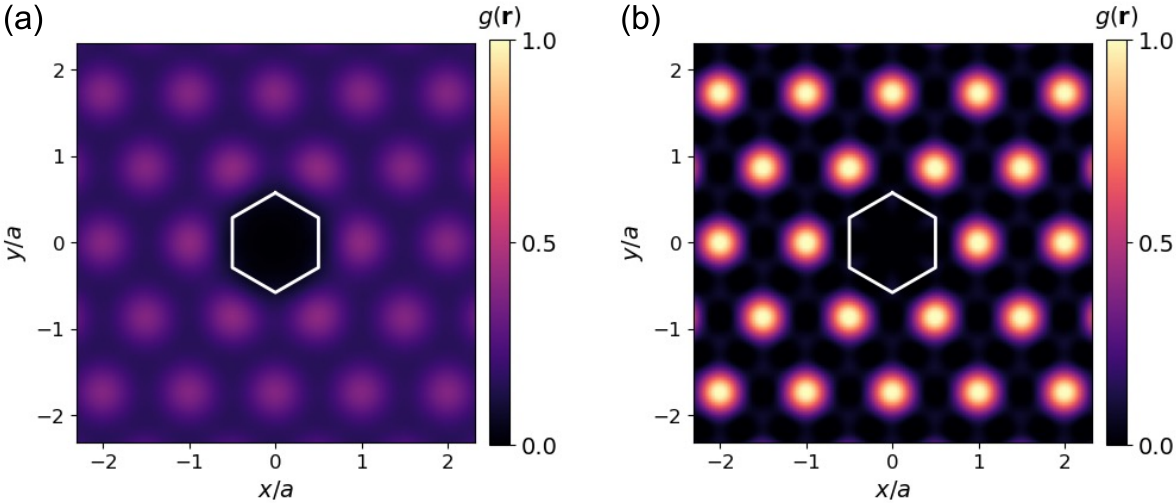}
    \caption{Pair correlation function $g(\r)$ for (a) $\mathcal K=0.4$ and (b) $\mathcal K=1.6$. The unit cell encloses three flux quanta.}
    \label{fig:pair_correlations}
\end{figure}

We characterize the real-space correlations utilizing the pair correlation function, which is defined as: 
\begin{equation}
    g(\r)=\frac{A}{N^2}\sum_{i\neq j}\langle\delta(\r-\r_i+\r_j)\rangle,
\end{equation}
where the average is performed over the many-body wavefunction $\Psi(\r_1,\cdots,\r_N)$ in Eq.~\eqref{psi_gstate}.
Fig.~\ref{fig:pair_correlations} shows the real-space pair correlation function. 
As shown in Figs.~\ref{fig:pair_correlations}(a)–\ref{fig:pair_correlations}(b), the pair correlation function $g(\r)$ evolves from weak, short-range modulations characteristic of the liquid-like fractionalized state to well-developed spatial modulations as $\mathcal K$ increases, signaling a transition toward a crystalline phase. 
More specifically, at small $r$, the correlation hole and short-range structure in Fig.~\ref{fig:pair_correlations}(a) closely resemble those of an FCI liquid.

\subsection{Many-body polarization}

To characterize the properties of the different many-body ground states, we compute the expectation value of the many-body polarization operator: 
\begin{equation}\label{app:MBpolarization}
     \zeta_j=\mel{\Psi_{\bm \phi}}{\exp\left(i\Delta\k_j\cdot\hat\R\right)}{\Psi_{\bm \phi}},
\end{equation}
where $j=1,2$, $\hat\R=\sum_{l=1}^{N}\hat\r_l$ is the many-body position operator and the momenta $\Delta\k_{1,2}$ are given by: 
\begin{equation}
    \Delta \k_1= -2\pi \frac{\mathbf z\times \L_2}{\L_1\times \L_2},\quad    \Delta \k_2= 2\pi \frac{\mathbf z\times \L_1}{\L_1\times \L_2}.
\end{equation}
The evolution of the phase of ${\zeta}_j$, $X_j=\Im\log\zeta_j/(2\pi)$, under flux threading determines the evolution of the center-of-mass position in units of $L_j$.  
In the FCI phase,
$X_1$ evolves with the threaded flux $\bm\phi=(0,\phi_2)$ and winds around the supercell over a flux-insertion cycle over the extended regime $0\leq\phi_2<6\pi$~\cite{Niu1985}.
In the crystalline phase, instead, $X_1$ no longer winds, and the trajectory of $\zeta_1$ forms a closed loop that does not enclose the origin.

Additionally, the absolute value of ${\zeta}_j$ quantifies the fluctuations of the many-body center-of-mass position. 
This can be readily understood by expanding the logarithm of the expectation value of Eq.~\eqref{app:MBpolarization} for small $|\Delta\k_j|$:
\begin{equation}
    -\log|\zeta_j|^2
    \approx
    \mel{\Psi_{\bm\phi}}
    {\left(\Delta\k_j\cdot\hat\R\right)^2
    }{\Psi_{\bm\phi}}
    -
    \mel{\Psi_{\bm\phi}}{
    \Delta\k_j\cdot\hat\R}{\Psi_{\bm\phi}}^2,
\end{equation}
which provides an estimate of the fluctuations of the many-body center of mass position. 
As shown in the main text, $|\zeta_1|$ approaches zero at the topological transition for a particular value of the threaded flux, causing $-\log|\zeta_1|$ to diverge and providing a sharp marker of the transition.

\subsection{System-size dependence of the plasma-to-dielectric transition}

\begin{figure}
    \centering
\includegraphics[width=0.4\linewidth]{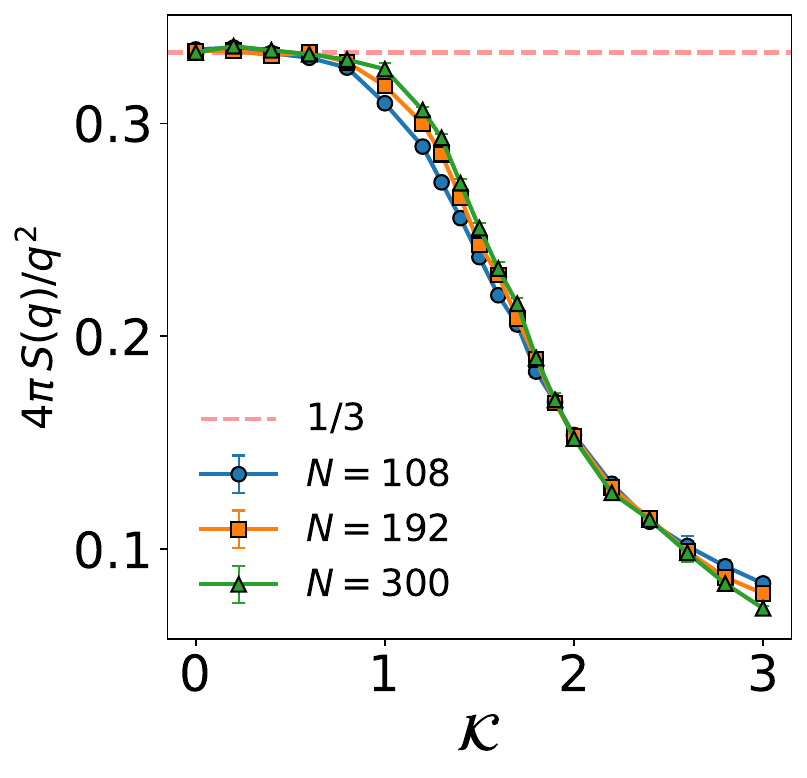}
\caption{The quadratic structure-factor coefficient $4\pi S(\q)/|\q|^2$ is shown as a function of the magnetic-field inhomogeneity $\mathcal K$, for $N= \{108,192,300 \}$. For $\mathcal K\lesssim1$, all sizes remain close to the perfect-screening value $1/3$ of the ideal FCI. At stronger modulation, the coefficient is suppressed, consistently signaling the onset of a finite dielectric response.}
    \label{fig:structure_factor_different_sizes}
\end{figure}

To assess finite-size effects in the structure-factor diagnostic, Fig.~\ref{fig:structure_factor_different_sizes} shows the quadratic coefficient of the structure factor $4\pi S(\q)/|\q|^2$ as a function of the magnetic-field inhomogeneity $\mathcal K$ for several system sizes. In the weak-modulation regime, $\mathcal K\lesssim1$, the results remain close to the perfect-screening value $1/3$, consistently identifying the FCI plasma. At stronger modulation, the coefficient decreases with $\mathcal K$, indicating the development of a finite  dielectric response. This behavior is common to all sizes and supports a transition to the dielectric phase. 

However, even at the relatively large system sizes that we consider, the transition region exhibits pronounced finite-size effects. The available sizes therefore do not permit a controlled finite-size collapse and thus a reliable extraction of the critical value $\mathcal K_{\rm c}$ from the structure factor. We consequently use these results to establish the evolution from perfect screening toward a dielectric response, while the BKT character of the transition follows from the Coulomb-gas renormalization-group analysis.

\section{One-flux model}\label{app:one_flux}

In this Appendix, we consider an emergent magnetic field in Eq.~\eqref{app:magnetic_field} where the modulation $\G_1$ and $\G_2$ corresponds to a unit cell that encloses one flux quantum $Q = 1$ and filling $\nu = 1/3$. Similarly to the $Q = 3$ model, in the weak-modulation regime, $Q = 1$ gives rise to a generalized Laughlin ground state.  We investigate the stability of the fractionalized ground state upon increasing the strength of the magnetic-field modulation. To this end, we examine the structure factor in the strong-modulation regime $\mathcal K\gg1$.

The limit $\mathcal K\to\infty$ is particularly instructive: if a crystalline instability is present, it should be most clearly manifested there. In this limit, physically, the electrons are forced to occupy a set of $3N$  discrete lattice sites, set by the minima of the emergent magnetic field. Thus, the many-body groundstate is described by Eq. \eqref{psi_gstate}, except that the electronic positions take values on a lattice. While this limit is obtained in our specific model of magnetic field by $\mathcal{K} \rightarrow \infty$, the obtained state is universal and independent of the shape of the emergent magnetic field. For example, in the solenoid model considered in Ref. \cite{Moitra2026}, it corresponds to the solenoid radius going to zero.  

To characterize the phase in this limit, we compute the structure factor performing Monte Carlo sampling. 
This limit is accessed in MC by restricting electrons to occupy a finite set of points corresponding to the magnetic field minima. 
This method was introduced in Ref.~\cite{iyer2026dispersionanyonblochbands} and realizes a lattice FCI state similar to the construction in Ref.~\cite{ji2026generalizedmodelfractionalquantum}.

\begin{figure}
    \centering
    \includegraphics[width=0.7\linewidth]{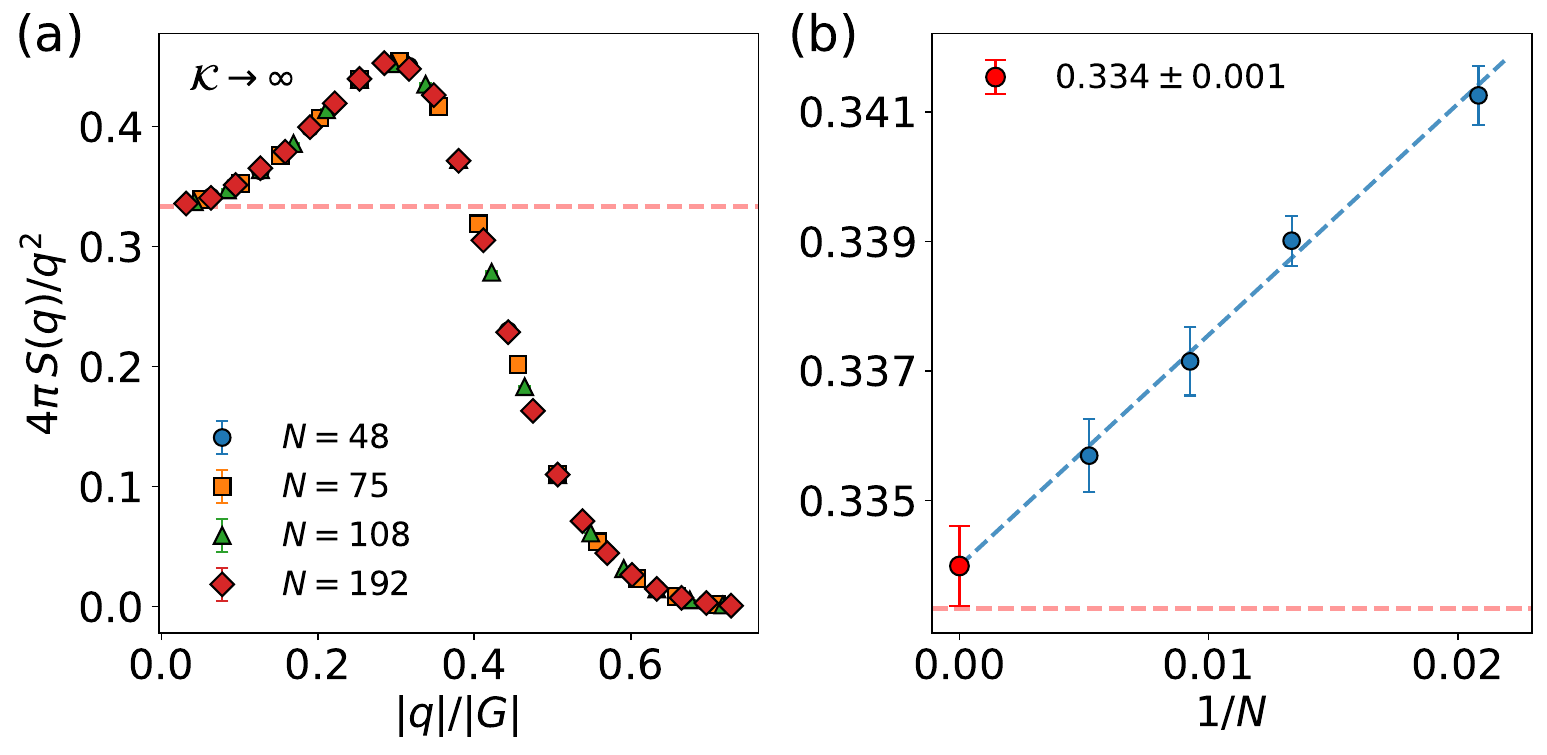}
    \caption{
    Structure factor in the one-flux model in the limit of an infinitely strong Kähler modulation. (a) $4\pi  S(q)/|q|^2$ as a function of $|q|/|G|$. The inset shows that this quantity approaches $|C| = 1/3$ as $q\rightarrow 0$, as expected for the FCI state. (b) The same quantity evaluated at the smallest available momentum, as a function of system size. Blue points denote Monte Carlo results for $N\in\{48,75,108,192\}$, while the red point shows the extrapolation to the thermodynamic limit.}
    \label{fig:one_flux_structure_factor}
\end{figure}

In Fig.~\ref{fig:one_flux_structure_factor}, we show the structure factor of the one-flux model for several system sizes in the infinite-$\mathcal K$ limit. 
The small-$\q$ quadratic coefficient remains finite and approaches the value expected for the fractional liquid, demonstrating that the system retains perfect screening and remains in the FCI phase even at infinite modulation. 
We find no evidence for a transition to a crystalline state.

The contrast with the $Q=3$ model has a simple physical origin: in the higher-winding (three-flux) model there is one magnetic field maximum per electron, so strong modulation locks each electron to a unit cell and produces a dielectic state. 
In the $Q=1$ model, there are three sites per electron, leaving extensive configurational freedom even at strong modulation, thus preserving the long-wavelength fractional-liquid correlations.

\section{Mapping to the neutral Coulomb gas model}\label{app_sec:neutral_gas}

The ideal FCI wave function on a torus in Eq. \eqref{psi_gstate} maps to a classical plasma Hamiltonian by the relation $|\Psi|^2=e^{-H}$ with
\begin{equation}
H=-2Q\sum_{i<j}V(z_i-z_j)
-2\sum_{\lambda=1}^{Q}V(Z-\alpha_\lambda)
+2\sum_i K(\boldsymbol r_i).
\label{eq:sm_plasma_hamiltonian}
\end{equation}
where $V(z) = \log|f(z)|$. We take $K(\boldsymbol r)$ to be lattice periodic, with one electron and $m$
flux quanta per unit cell.  In the infinitely strong-modulation limit of $K(\r)$, the
electrons are pinned near a set of lattice sites
$\mathcal R=\{R_1,\ldots,R_{N_e}\}$ given by the minima of $K(\r)$. We note that this is true independently of the form of $K(\r)$ with the only assumption that there is exactly one maximum per unit cell. This frozen configuration is used as a
reference state; particle-hole pairs have a non-infinite cost only at finite modulation.

We create $n$ neutral particle--hole pairs by removing electrons from sites
$h_a\in\mathcal R$ and placing them at $p_a=u_a+\delta_a$, where $u_a \in \mathcal R$ is an
occupied reference site and $\delta_a$ lies within its microscopic core.
Let $\mathcal U=\mathcal R\setminus\{h_1,\ldots,h_n\}$.  Subtracting the
reference energy eliminates all interactions involving only sites in
$\mathcal U$.  It is useful to define
\begin{equation}
\Phi_0(\delta)=\sum_{R\in\mathcal R}V(\delta-R),
\qquad
\mu_0=\sum_{R\in\mathcal R\setminus\{0\}}V(R).
\label{eq:sm_reference_potentials}
\end{equation}
Translation invariance makes these quantities independent of the chosen
reference cell.  The identities
\begin{align}
\sum_{a,u\in\mathcal U}V(p_a-u)
&=\sum_a\Phi_0(\delta_a)-\sum_{a,b}V(p_a-h_b),\\
\sum_{a,u\in\mathcal U}V(h_a-u)
&=n\mu_0-2\sum_{a<b}V(h_a-h_b)
\end{align}
then give
\begin{align}
\Delta H={}&-2Q\left[
\sum_{a<b}V(p_a-p_b)+\sum_{a<b}V(h_a-h_b)
-\sum_{a,b}V(p_a-h_b)\right]+\sum_a\epsilon_{\rm c}(\delta_a)+\Delta H_{\rm CM},
\label{eq:sm_defect_energy}\\
\epsilon_{\rm c}(\delta)={}&
-2QV(\delta)
+2\bigl[K(\delta)-K(0)\bigr].
\label{eq:sm_core_energy}
\end{align}
where $\Phi_0(\delta)+\mu_0 = V(\delta)$ can be shown using the torus PBC. Thus, $\epsilon_{\rm c}$ is the local cost of creating a particle--hole pair,
which comprises the interaction of the particle with its neighbor in the microscopic core and the different in the K\"ahler potential between the positions of the particle and the hole. 
The centre-of-mass term $\Delta H_{\rm CM}$ produces a global term which, to leading order, is $O(\Delta Z/|L_1|)$ with $\Delta Z = \sum_a(p_a-h_a)$. In the thermodynamic limit, the latter quantity vanishes away from the isolated COM zeros. More generally, it remains nonextensive and generates no logarithmic defect interaction; it therefore does not affect the bulk Coulomb-gas RG and we drop it in the following.

Introduce $2n$ charges $q_\mu=+1$ at $x_\mu=p_\mu$ and $q_\mu=-1$ at
$x_\mu=h_{\mu-n}$, so that $\sum_\mu q_\mu=0$.  Equation
\eqref{eq:sm_defect_energy} becomes
\begin{equation}
\Delta H=-2Q\sum_{\mu<\nu}q_\mu q_\nu V(x_\mu-x_\nu)
+\sum_{a=1}^{n}\epsilon_{\rm c}(\delta_a).
\label{eq:sm_neutral_gas}
\end{equation}
For separations $r\gg |\delta_a|$, the interaction cannot resolve the
intra-core displacement, and $p_a$ may be replaced by the coarse-grained
position $u_a$.  All microscopic information is then contained in the
dimensionless pair fugacity
\begin{equation}
y^2\equiv Y=
\int_{\mathcal C}\frac{d^2\delta}{a^2}
e^{-\epsilon_{\rm c}(\delta)},
\label{eq:sm_fugacity}
\end{equation}
where $\mathcal C$ is the core area of the particle-defect and $a$ is a microscopic cutoff of order
the core.  A change of cutoff only redefines the bare value
of $y$.  In the dilute limit, different cores do not overlap and their
integrals factorize.  The grand-canonical defect partition function is
therefore
\begin{equation}
\frac{\mathcal Z}{\mathcal Z_0}=
\sum_{n=0}^{\infty}\frac{y^{2n}}{(n!)^2}
\prod_{\mu=1}^{2n}\int\frac{d^2s_\mu}{a^2}
\exp\!\left[
\frac{2m}{\epsilon}\sum_{\mu<\nu}q_\mu q_\nu V(s_\mu-s_\nu)
\right],
\qquad \sum_\mu q_\mu=0.
\label{eq:sm_partition_function}
\end{equation}
At distances $a\ll r\ll |L_{1,2}|$,
$V(r)=\log(r/a)+\mathrm{const.}+O(r^2/|L_{1,2}|^2)$; the constant can be
absorbed into $y$.  Equation~\eqref{eq:sm_partition_function} thus reduces to
the standard neutral two-dimensional Coulomb gas.  Defining a running coupling
$Q\epsilon^{-1}$ with $Q\epsilon^{-1}(0)=Q$, its leading-order RG equations are \cite{drouintouchette2022kosterlitzthoulessphasetransitionintroduction}
\begin{align}
\frac{d\epsilon^{-1}}{d\ell}&=-4Q\pi^2\epsilon^{-2}y^2+O(y^4),\\
\frac{dy}{d\ell}&=(2-Q\epsilon^{-1})y+O(y^3).
\label{eq:sm_rg}
\end{align}
For small fugacity and $Q\epsilon^{-1}>2$, particle--hole defects remain confined, giving a dielectric phase.  
Upon increasing $y$, the flow crosses the BKT separatrix and defects proliferate forming a perfectly screening plasma. 
The endpoint of the critical separatrix at $Q\epsilon^{-1}=2$ gives the universal Nelson-Kosterlitz jump of $\epsilon^{-1}$ from $\epsilon^{-1} = 2/Q$ on the dielectric side to $\epsilon^{-1} = 0$ on the plasma side.  The detailed form and amplitude of
$K(\boldsymbol r)$ determine the fugacity through
Eq.~\eqref{eq:sm_fugacity}, but not the long-distance Coulomb-gas interaction,
the RG structure, or the universal jump.

\end{document}